\documentstyle[12pt]{article}
\def\sect#1{\section*{#1}\addtocounter{section}{1}\setcounter{equation}{0}}
\def\emline#1#2#3#4#5#6{%
       \put(#1,#2){\special{em:moveto}}%
       \put(#4,#5){\special{em:lineto}}}

\begin{document}
\pagestyle{empty}
\begin{flushright}
KAIST/THP-96/701
\end{flushright}
\begin{center}
{\Large {\bf Root Systems and Boundary Bootstrap} }
\end{center}
\begin{center}
{\bf J. D. Kim\footnote{jdkim@sorak.kaist.ac.kr}}  \\
\vspace{0.1cm}
{\it Department of Physics \\
KAIST, Taejon 305-701 Korea} \\
\vspace{0.3cm}
{\bf Y. Yoon\footnote{cem@hepth.hanyang.ac.kr} } \\
\vspace{0.1cm}
{\it Department of Physics    \\
Hanyang University, Seoul 133-791 Korea}
\end{center}
\vspace{2.0cm}

\begin{abstract}
The principle of boundary bootstrap plays a significant role
in the algebraic study of the purely elastic boundary reflection matrix
$K_a(\theta)$ for integrable quantum field theory defined on a space-time
with a boundary. However, general structure of that principle in the
form as was originally introduced by Fring and K\"oberle has remained unclear.
In terms of a new matrix
$J_a(\theta)=\sqrt{K_a(\theta)/K_{\bar{a}}(i\pi +\theta)}$,
the boundary bootstrap takes a simple form.
Incidentally, a hypothesised expression of the boundary reflection matrix
for simply-laced $ADE$ affine Toda field theory defined on a half line
with the Neumann boundary condition is obtained in terms of geometrical
quantities of root systems \`a la Dorey.
\end{abstract}

\newpage
\pagestyle{plain}

\sect{1. Introduction}
The boundary reflection matrix in quantum field theory is conceived
to describe a reflection process of particles against a (spatial) boundary
of a space-time. When the property of integrability of the quantum field
theory defined on a whole line continues to hold even in the presence
of a boundary, consistency requirements such as the principle of boundary
bootstrap strongly constrain the exact boundary reflection matrix.

A significant step toward the algebraic study of the exact boundary
reflection matrix was put forward by introducing the boundary Yang-Baxter
equation\cite{Che} about a decade ago. However, in case of non-diagonal
reflection, this equation alone is not sufficient to allow one to find
the scalar function of the boundary reflection matrix.
About ten years had passed until the scalar function
was finally found by introducing the crossing-unitarity relation\cite{GZ}.

In case of diagonal reflection where types of particles do not change,
the boundary Yang-Baxter equation becomes trivially satisfied. So one
need another condition like the boundary bootstrap equation\cite{FK}.
Subsequently, a variety of solutions of the algebraic equations
for affine Toda field theory(ATFT) has been constructed
explicitly\cite{GZ,FK,Sasaki,CDRS,CDR}.
However, proper interpretations to these solutions in the framework of
Lagrangian quantum field theory was not given\footnote{There are some works
which aim to relate parameters appearing in the boundary potential and
formal parameters arising from solutions of the algebraic
equations; for the sine-Gordon theory at a generic point in semi-classical
analysis\cite{SSW} and at the free fermion point\cite{AKL} where one may
use the method\cite{GZ} of mode expansion for the field as an operator.}.

In order to have a direct access to the boundary reflection matrix,
perturbative approach has been developed and a quantum field theoretic
definition of the boundary reflection matrix was proposed\cite{Kim}.
A complete set of conjectures for the exact boundary reflection matrix
for simply-laced $ADE$ affine Toda field theory defined on a half line
with the Neumann boundary condition was obtained\cite{Kim2}.
It is noted that each of the solutions is not \lq minimal' among all
possible solutions of the algebraic equations in the usual
sense of the total number of poles and zeros on the physical strip.

General structure of the boundary bootstrap in the form as was originally
introduced by Fring and K\"oberle has remained unclear\cite{FK}.
In this letter, a new matrix
$J_a(\theta)=\sqrt{K_a(\theta)/K_{\bar{a}}(i\pi +\theta)}$
which renders the boundary bootstrap more tractable is introduced
and a hypothesised expression of the boundary reflection matrix
for simply-laced $ADE$ affine Toda field theory defined on a half line
with the Neumann boundary condition is obtained in terms of
geometrical quantities of root systems \`a la Dorey.

In section 2, the geometry associated with the Coxeter element of the Weyl
group of root systems which also played an important role in
the geometric formulation of the $S$-matrix\cite{Dorey,FO}
is briefly reviewed to set up a notation. In section 3, the new form of
the boundary bootstrap is introduced in terms of $J_a(\theta)$ function
and some discussions on its properties are given. Then, the hypothesised
expression of the boundary reflection matrix for simply-laced $ADE$ affine
Toda field theory defined on a half line with the Neumann boundary
condition is obtained in terms of geometrical quantities of root systems.
Finally some conclusions are made in section 4.

\sect{2. Geometry of root systems}
The geometric expression of the exact $S$-matrix for
simply-laced $ADE$ affine Toda field theory can be written in various
ways depending on a chosen set of representatives of the Weyl
orbits. The notation of ref.\cite{Corrigan} will be taken here.

Let simple roots $\alpha_i~(i=1,\ldots,n)$ for a simply-laced Lie algebra
$g$ with rank $n$ be divided into two sets such that
the roots in each set are mutually orthogonal. The members of the
two sets are distinguished by assigning a colour to them, either black
or white. Let the simple roots be labelled so that
\begin{equation}
\bullet =\{1,2,\ldots,k\},~~~~~~ \circ=\{k+1,k+2,\ldots,n\}.
\end{equation}

For any root $x$, an elementary Weyl reflection $w_i$ corresponding
to the simple root $\alpha_i$ is defined by
\begin{equation}
w_i( x )=x-2 \frac{ x \cdot \alpha_i }{ \alpha_i^2 } \alpha_i.
\end{equation}
The Weyl group is generated by these elementary Weyl reflections.
A Coxeter element of the Weyl group is a product over the simple roots
of the elementary Weyl reflections in any choice of ordering.
With the above choice of ordering of the simple roots,
the Coxeter element is defined by
\begin{equation}
w=w_{\bullet} w_{\circ} = w_1 \ldots w_k w_{k+1} \ldots w_n.
\end{equation}
The order of the Coxeter element is $h$, the Coxeter number.

Finally, root vectors $\phi_i ~(i=1,\ldots,n)$ are selected
as representatives of the Weyl orbits as follows:
\begin{equation}
\phi_i=w_n w_{n-1} \ldots w_{i+1}(\alpha_i).
\end{equation}
Then, the image of each $\phi_i ~(i=1,\ldots,n)$ under the inverse
Coxeter element is a positive root and successive images remain positive
until they all change sign, remaining
negative subsequently for the rest of the orbit.

With this machinery, the hypothesised expression of the exact $S$-matrix
for simply-laced $ADE$ affine Toda field theory is written
in the following form:
\begin{equation}
S_{ab}(\theta)=\prod_{p=0}^{h-1} \{ 2p+1+\epsilon_{ab} \}^{1/2 (\lambda_a
\cdot w^{-p} \phi_b) },
\label{S}
\end{equation}
where
\begin{equation}
 \{ x \} = \frac{(x-1) (x+1)} {(x-1+2 B) (x+1-2 B)},
 ~~~~ (x) = \frac{sh(\theta /2+i \pi x /2 h)} {sh(\theta /2-i \pi x /2 h)}.
\label{Block}
\end{equation}
$\theta=\theta_a-\theta_b$ is the difference of the rapidities and
$\lambda_a$ are dual vectors such that
$(\lambda_i \cdot \alpha_j)=\delta_{ij}$.
The coupling dependence enters through the universal function
$B(\beta)=\beta^2/(\beta^2 + 4\pi)$.
$\epsilon_{ab}$ is defined as follows depending on the colour
of the pair $a,b$:
\begin{equation}
\epsilon_{\bullet\bullet}= \epsilon_{\circ\circ}=0,
~~~~~~ \epsilon_{\circ\bullet}= -\epsilon_{\bullet\circ}=1.
\end{equation}

\sect{3. Boundary bootstrap}
It is usually supposed that particles on a half line
$(-\infty <x \leq 0)$ scatter as if the boundary were absent.
In other words, scattering of particles on a half line can be
described by the same $S$-matrix of the system defined on a whole line.

In the algebraic approach, the boundary reflection matrix
is defined in terms of Zamolodchikov-Faddeev algebra\cite{GZ}:
\begin{equation}
A_a ^{\dagger}(\theta) B ~=~ K_a (\theta) A_a^{\dagger}(-\theta) B,
\label{K}
\end{equation}
where $A_{a}^{\dagger}$ is the creation operator of particle $a$
and $B$ is the boundary creation operator.
Consistency requirements of the boundary reflection with
the scattering and the three-point vertex function lead
to some algebraic relations which impose stringent constraints
to the solution of the boundary reflection matrix.

To begin with, the general unitarity requirement leads to the boundary
unitarity relation:
\begin{equation}
K_a(\theta)K_a(-\theta)=1.
\label{BU}
\end{equation}
Consistency requirement on the two particle process yields a constraint
which is called the boundary Yang-Baxter equation:
\begin{equation}
K_b(\theta_b) S_{ab}(\theta_a +\theta_b) K_a(\theta_a)
S_{ab}(\theta_a -\theta_b) =
S_{ab}(\theta_a -\theta_b) K_a(\theta_a) S_{ab}(\theta_a +\theta_b)
K_b(\theta_b).
\label{BYB}
\end{equation}
When types of particles do not change during the reflection process
as in the case of ATFT,
eq.(\ref{BYB}) is automatically satisfied. Crossing-unitarity relation is
\begin{equation}
K_a(\theta) K_{\bar a}(i \pi +\theta) = S_{aa}(2 \theta).
\label{BCU}
\end{equation}
This relation is non-linear in $K$, which is effective particularly
in solving the scalar function of non-diagonal boundary reflection matrix.

Consistency requirement of the boundary reflection with the three-point vertex
function leads to the boundary bootstrap equation:
\begin{equation}
K_c(\theta) (-i f_c^{ab}) = (-i f_c^{ab}) K_a(\theta+i {\bar \theta_{ac}^{b}})
S_{ab}(2 \theta+i {\bar \theta_{ac}^{b}} -i {\bar \theta_{bc}^{a}} ) 
K_b(\theta-i {\bar \theta_{bc}^{a}} ).
\label{BB}
\end{equation}
$f_{c}^{ab}$ is the three-point vertex function. The fusing angles
$\theta_{ab}^{c}$ is defined as $m_c^2=m_a^2+m_b^2-2m_a m_b
\cos \theta_{ab}^{c}$ and ${\bar \theta}= i\pi-\theta$.
eq.(\ref{BB}) is represented pictorially in Fig. 1.
For diagonal reflection, only the crossing-unitarity relation and
the boundary bootstrap equation produce non-trivial constraints
for the boundary reflection matrix.

\unitlength=1.00mm
\special{em:linewidth 0.5pt}
\linethickness{0.5pt}
\begin{picture}(120.00,57.00)(2,0)
\emline{50.00}{10.00}{1}{50.00}{50.00}{2}
\emline{120.00}{10.00}{3}{120.00}{50.00}{4}
\emline{35.00}{10.00}{5}{50.00}{25.00}{6}
\emline{50.00}{25.00}{7}{45.00}{30.00}{8}
\emline{45.00}{30.00}{9}{40.00}{45.00}{10}
\emline{45.00}{30.00}{11}{33.00}{38.00}{12}
\put(35.00,10.00){\vector(1,1){8.00}}
\put(45.00,30.00){\vector(-3,2){8.00}}
\put(45.00,30.00){\vector(-1,3){2.67}}
\put(28.00,08.00){\makebox(0,0)[cc]{$c$}}
\put(28.00,38.00){\makebox(0,0)[cc]{$a$}}
\put(34.00,48.00){\makebox(0,0)[cc]{$b$}}
\emline{105.00}{10.00}{13}{113.00}{18.00}{14}
\emline{113.00}{18.00}{15}{120.00}{20.00}{16}
\emline{120.00}{20.00}{17}{100.00}{30.00}{18}
\emline{113.00}{18.00}{19}{120.00}{37.00}{20}
\emline{120.00}{37.00}{21}{113.00}{50.00}{22}
\put(99.00,07.00){\makebox(0,0)[cc]{$c$}}
\put(94.00,30.00){\makebox(0,0)[cc]{$a$}}
\put(106.00,50.00){\makebox(0,0)[cc]{$b$}}
\put(75.00,30.00){\makebox(0,0)[cc]{=}}
\put(105.00,10.00){\vector(1,1){5.00}}
\put(120.00,20.00){\vector(-2,1){10.00}}
\put(120.00,37.00){\vector(-1,2){3.67}}
\put(49.50,21.50){\oval(5.00,7.00)[lb]}
\put(46.00,14.00){\makebox(0,0)[cc]{$\theta$}}
\put(75.00,02.00){\makebox(0,0)[cc]{Figure 1.}}
\end{picture}

The above boundary bootstrap equation involves the $S$-matrix
in a rather non-trivial way.
This fact makes it difficult to see the general structure of it.
In order to separate the $S$-matrix from the boundary bootstrap equation,
let the arguement $\theta$ be shifted by an amount of $i\pi$
and take the conjugation of the particle indices. Then, it yields
\begin{equation}
K_{\bar c}(\theta+i \pi)=K_{\bar a}(\theta+i \pi+i {\bar \theta_{ac}^{b}})
S_{ab}(2\theta+i {\bar \theta_{ac}^{b}} -i {\bar \theta_{bc}^{a}} ) 
K_{\bar b}(\theta + i \pi-i {\bar \theta_{bc}^{a}} ).
\label{SBB}
\end{equation}
Here the facts that $S(\theta)$ is $2\pi i$-periodic and
$S_{\bar{a}\bar{b}}(\theta) = S_{ab}(\theta)$ are used.
Fusing angle $\theta_{ab}^c$ does not change under the conjugation.

Now, it seems natural to introduce a new function $J_a(\theta)$:
\begin{equation}
J_a(\theta)=\sqrt{K_a(\theta)/K_{\bar{a}}(i\pi +\theta)}
=K_a(\theta)/\sqrt{S_{aa}(2 \theta)}.
\label{J}
\end{equation}
The second equality follows from the crossing-unitarity relation.
On replacing the definition of $J_a(\theta)$ into eq.(\ref{BB})
divided by eq.(\ref{SBB}), one gets a very simple equation
for the boundary bootstrap:
\begin{equation}
J_c(\theta)=
J_a(\theta+i {\bar \theta_{ac}^{b}}) J_b(\theta -i {\bar \theta_{bc}^{a}} ).
\label{NBB}
\end{equation}
This equation may be depicted as in Fig. 2. At first glance,
this equation seems to have nothing to do with the boundary conditions.
But, in fact it does have something!
For instance, in order to solve eq.(\ref{NBB}) one needs to know
the limiting behavior of $J_a(\theta)$ at the strong and weak couplings,
which depends on a given boundary potential.

There is one interesting coincidence. Namely, the same form of the
equation as eq.(\ref{NBB}) already appeared in ref.\cite{CDR},
where $J_a(\theta)$ is interpreted as the classical
limit (where $S(\theta)$ tends to unity) of the exact $K_a(\theta)$.
In general, the classical limit of $K_a(\theta)$ need not be unity,
as in cases with integrable non-trivial boundary potentials.
For the present case of the Neumann boundary condition,
$J_a(\theta)$ also tends to unity as $\beta \rightarrow 0$.

\begin{picture}(60.00,62.00)(-40,-5)
\emline{10.00}{20.00}{1}{35.00}{35.00}{2}
\emline{35.00}{35.00}{3}{26.00}{06.00}{4}
\emline{35.00}{35.00}{5}{49.00}{52.00}{6}
\put(21.80,27.00){\vector(3,2){1.50}}
\put(30.30,19.80){\vector(1,3){1.00}}
\put(5.00,20.00){\makebox(0,0)[lb]{$a$}}
\put(30.00,04.00){\makebox(0,0)[lb]{$b$}}
\put(50.00,47.00){\makebox(0,0)[lb]{$c$}}
\put(27.00,40.00){\makebox(0,0)[lb]{$\theta_{ac}^b$}}
\put(37.20,33.00){\oval(11.80,9.10)[lt]}
\put(38.50,27.50){\makebox(0,0)[lb]{$\theta_{bc}^a$}}
\put(34.00,37.50){\oval(6.80,12.50)[rb]}
\put(25.00,-3.00){\makebox(0,0)[lb]{Figure 2.}}
\end{picture}

In terms of $J_a(\theta)$, the unitarity relation eq.(\ref{BU}) changes into
\begin{equation}
J_a(\theta)J_a(-\theta)=1,
\label{NBU}
\end{equation}
and the crossing-unitarity relation eq.(\ref{BCU}) simplifies to
\begin{equation}
J_a(\theta) J_{\bar a}(i \pi +\theta) = 1.
\label{NBCU}
\end{equation}

With the known conjecture\cite{Kim2} of the boundary reflection matrix
for simply-laced $ADE$ affine Toda field theory defined on a half line
with the Neumann boundary condition, it is not a difficult observation
to derive the following hypothesis:
\begin{equation}
J_{b}(\theta)=\prod_{p=0}^{h-1} \left[ 2p+1/2+\epsilon_{b}
\right]^{1/2 \sum_{a} (\lambda_a \cdot w^{-p} \phi_b)},
\label{JJ}
\end{equation}
where
\begin{equation}
 [ x ] = \frac{(x-1/2) (x+1/2)} {(x-1/2+B) (x+1/2-B)}.
\label{HBlock}
\end{equation}
$\theta$ is the rapidity
and $\epsilon_b$ is defined as follows depending on the colour of $b$:
\begin{equation}
\epsilon_{\bullet}=1,~~~~~~\epsilon_{\circ}=0.
\end{equation}
For reader's reference, some identities are listed below:
\begin{equation}
\{x\}_{2\theta}=[x/2]_{\theta}/[h-x/2]_{\theta},~~~[2h+x]=[x],~~~[-x]=1/[x].
\end{equation}
The above formula for the $J$-matrix in eq.(\ref{JJ}) has a very similar
dependence on the Coxeter element as the one for the $S$-matrix
in eq.(\ref{S}) and can be shown to satisfy all the necessary algebraic
constraints quite analogously as in ref.\cite{Dorey}
with minor modifications.

\sect{4. Conclusions}
For purely elastic boundary reflection, the principle of boundary bootstrap
plays a significant role in the algebraic study on the exact
boundary reflection matrix
since the boundary Yang-Baxter equation becomes trivially satisfied.

The boundary bootstrap in its original form has not allowed
an easy attack on its general structure\cite{FK}.
In this letter, the new matrix
$J_a(\theta)=\sqrt{K_a(\theta)/K_{\bar{a}}(i\pi +\theta)}$
which renders the boundary bootstrap more tractable was introduced
and then by analysing the known conjectures on the boundary reflection
matrix with the new version of the boundary bootstrap,
the hypothesised expression of the boundary reflection matrix
for simply-laced $ADE$ affine Toda field theory defined on a half line
with the Neumann boundary condition was obtained in terms of root systems.

Boundary conditions which are compatible with classical or quantum
integrability has been investigated for ATFTs associated with simply-laced
Lie algebras as well as non-simply-laced Lie
algebras\cite{GZ,CDRS,CDR,BCDR,PZ,PRZ}.
Classical boundary reflection matrices corresponding to the
various choices of the integrable boundary condition have been constructed
by linearising the equation of motion around a background solution in
refs.\cite{CDRS,CDR}, where some conjectures on the exact boundary reflection
matrices have been also made.

Further studies on the new version of the boundary bootstrap
would be useful for finding exact boundary reflection matrices
corresponding to integrable non-trivial boundary potentials.
To this end, one should take into account renormalisation of
the boundary parameters\cite{PZ,PRZ}
which is related to the limiting behavior of boundary reflection matrix
at the strong and weak couplings and stability of the particle
spectrum\cite{FS}.

It is also hoped that deeper understandings in boundary reflection matrix 
will provide a new sort of insights into the unified formulation of ATFTs
associated with simply-laced Lie algebras as well as non-simply-laced Lie
algebras. Boundary reflection matrices
for some non-simply-laced ATFTs have been obtained in refs.\cite{Sasaki,Kim3}.

\section*{Acknowledgement}
One(JDK) of the authors wishes to thank the department of physics
in Hanyang University for warm hospitality.
We would like to thank Sang Jin Sin and Bum-Hoon Lee for 
stimulating discussions.
This work is supported in part by the Ministry of Education through 
Grant No. BSRI-2441 and Center for Theoretical Physics in SNU.

\end{document}